\documentclass[preprint2]{aastex6}

\usepackage{graphicx,epsfig,fancyhdr,psfig,rotating,amsmath,epsf,txfonts,natbib,epstopdf,multirow}
\def \kms {{\rm km\;s$^{-1}$}}

\def \arcsec {$^{''}$}
\def \siiv {Si\,{\sc iv}}
\def \cii {C\,{\sc ii}}

\begin{document}
\title{Magnetic loops above a small flux-emerging region observed by IRIS, Hinode and SDO}
\author{
Zhenghua Huang
}
\affil{Shandong Provincial Key Laboratory of Optical Astronomy and Solar-Terrestrial Environment, Institute of Space Sciences, Shandong University, Weihai, 264209 Shandong, China; {\it z.huang@sdu.edu.cn}}

\begin{abstract}
I report on observations of a set of magnetic loops above a region with late-phase flux emergence taken by IRIS, Hinode and SDO.
The loop system consists of many transition region loop threads with size of 5--12\arcsec\ in length and $\sim0.5$\arcsec in width and coronal loops with similar length and $\sim2$\arcsec width.
Although the loop system consists of threads with different temperatures,
most individual loop thread have temperature in a narrow range.
In the middle of the loop system, it shows clear systematic blue-shifts of about 10\,\kms\ in the transition region that is consistent with a flux emerging picture,
while red-shifts of about 10\,\kms\ in the corona is observed.
The nonthermal velocity of the loop system are smaller than the surrounding region in the transition region but are comparable in the corona.
The electron densities of the coronal counterpart of the loop system range from  $1\times10^9$\,cm$^{-3}$ to $4\times10^9$\,cm$^{-3}$.
Electron density of a transition region loop is also measured and found to be about $5\times10^{10}$\,cm$^{-3}$, a magnitude larger than that in the coronal loops.
In agreement with imaging data, the temperature profiles derived from the differential emission measurement technique confirms that some of the loops have been heated to corona.
Our observations indicate that the flux emergence in its late phase is much different from that at the early stage.
While the observed transition region is dominated by emerging flux,
these emerging loops could be heated to corona and the heatings most likely take place only after they reaching the transition region or lower corona.
\end{abstract}

\maketitle

\section{Introduction}
\label{sect_intro}
The ionised gas in the solar atmosphere could be highly structured by the magnetic field,
forming a variety of loop features therein.
These loop features, namely magnetic loops, are one of the fundamental building block of the solar atmosphere.
These loops are normally brighter than the background and could be easily traced in the remote-sensing data. Therefore,
they are popular objects used to investigate transporting process of magnetic flux and energy from the solar interior to the outer solar atmosphere.

\par
The physical parameters, such as velocity, density and temperature of magnetic loops, are directly related to the pressure and emissivity, whose distribution along the loop length depends on heating therein\,\citep[e.g.][]{1998Natur.393..545P,2006SoPh..234...41K,2007ApJ...664.1214P,2008ApJ...677.1395W,2008ApJ...682.1351K,2013ApJ...771..115V,2014LRSP...11....4R,2018ApJ...856..178P}. 
Therefore, these parameters are crucial to understand the mechanism and spatial distribution of heating in the loop.
Measurements of those parameters have been achieved, but mostly for magnetic loops in the corona thanks to their relatively stable geometries\,\citep[an overview of these observations of coronal loops can be found in][]{2017ApJ...842...38X}.
However, it is challenging to obtain such measurements for cool transition region loops, 
because this class of loops are highly dynamic on time scale of minutes\,\citep{1998SoPh..182...73K,2015ApJ...810...46H}.
In the \textit{SOHO} era, magnetic loops with line-of-sight velocities from a few tens to a hundred kilometres per second have been reported in \textit{CDS} O\,{\sc V} ($2.5\times10^5$\,K) observations\,\citep{1997SoPh..175..511B,1998SoPh..182...73K, 2003A&A...406..323D} and \textit{SUMER} O\,{\sc VI} ($3.2\times10^5$\,K) data\,\citep{2000ApJ...533..535C,2004A&A...427.1065T,2006A&A...452.1075D}.
Since the Interface Region Imaging Spectragraph\,\citep[IRIS,][]{2014SoPh..tmp...25D} achieved its first light in 2013, the transition region has been observed with unprecedented spatial and temporal resolutions.
With IRIS observations, \citet{2015ApJ...810...46H} confirmed that this class of loops are diverse and dynamic,
and they observed siphon flows of 10--20\,\kms\ in a group of loops.
Regarding density and temperature, accurate measurements for cool transition region region loops are rare due to lack of  suitable spectroscopic data.

\par
Dynamic phenomena in cool transition region loops could also provide insight for the energy and mass transportation therein.
It has reported that interaction between the cool transition region loops could produce explosive events\,\citep{2015ApJ...810...46H,2017MNRAS.464.1753H,2018ApJ...854...80H}, which are signature of energy releases in the solar transition region via magnetic reconnection\,\citep[e.g.][etc.]{1991JGR....96.9399D,1997Natur.386..811I,2014ApJ...797...88H,2018arXiv180610205L}.
It has also observed that interaction between cool transition region loops could result in heating and then forming hotter coronal loops\,\citep[e.g.][]{2017A&A...603A..95A,2018A&A...611A..49Y,2018arXiv180611045C}.

\par
Here, I report on multi-wavelength observations of a set of magnetic loops above a region with flux emergence at its late phase.
Rather than an individual loop, I will focus on the global observational characteristics of these loops.
I will investigate the dynamics of the loops and its thermal structures coupling from chromospheric to coronal temperatures.
By analysing the spectroscopic data, I will also determine the physical parameters of the loops, such as velocities, densities and temperatures.
Basing on these observations, I will have discussion on behaviors of flux emergence at its late phase and possible heating processes of these emerging loops.
The study is organised as followed: the data information is described in Section\,\ref{sect_data}; the analysis results and discussion are given in Section\,\ref{sect_results}; the conclusion is given in Section\,\ref{sect_conclusion}.

\section{Observations and data reductions}
\label{sect_data}
\subsection{Data description}
The data were acquired during an observing campaign in coordination with the Goode Solar Telescope\,\citep[GST,][]{2003JKAS...36S.125G,2010AN....331..620G,2010AN....331..636C} operated in the Big Bear Solar Observatory\,\citep[BBSO,][]{1970S&T....39..215Z}, the Interface Region Imaging Spectragraph\,\citep[IRIS,][]{2014SoPh..tmp...25D}, Hinode\,\citep{2007SoPh..243....3K} and the Solar Dynamics Observatory\,\citep[SDO,][]{2012SoPh..275....3P}.
On August 26 2017, 
the region of interest was observed by a raster scan of IRIS spectrograph from 18:00\,UT to 19:27\,UT, and also by Hinode instruments.
During this period of time, the GST did not target on this region due to the other arrangement.

\par
Here, I analysed the spectroscopic data obtained by the EUV Imaging Spectrometer \,\citep[EIS,][]{2007SoPh..243...19C} aboard Hinode and IRIS,  UV and EUV images taken with the Atmospheric Imaging Assembly\,\citep[AIA,][]{2012SoPh..275...17L} aboard SDO and IRIS, soft X-ray images taken with the X-Ray Telescope\,\citep[XRT,][]{2007SoPh..243...63G} aboard Hinode and line-of-sight magnetograms measured by the Helioseismic and Magnetic Imager\,\citep[HMI,][]{2012SoPh..275..207S} aboard SDO.

\begin{table}[!ht]
\centering
\caption{Summary of the imaging data analysed presently.}
\label{tab_dimg}
\begin{tabular}{c c c c c}
\hline
Inst. & passband & cadence & pixel size&Peak Tem- \\
 &(\AA)& (s) &   & perature (K)\\
\hline
\hline
IRIS$^\dag$ &1330&65&0.17\arcsec$\times$0.17\arcsec&$\sim2.5\times10^4$\\
  &1400&---&---&$\sim8\times10^4$\\
  &2796&---&---&$\sim1\times10^4$\\
\hline
AIA &1700&24&0.6\arcsec$\times$0.6\arcsec&$5\times10^3$\\
&304&12&---&$5\times10^4$\\
&171&---&---&$6.3\times10^5$\\
&193&---&---&$1.3\times10^6$\\
&94&---&---&$6.3\times10^6$\\
\hline
XRT &Al\_poly&$\sim40$&1\arcsec$\times$1\arcsec&$8\times10^6$\\
\hline
HMI &magnetogram&45$^{*}$&0.5\arcsec$\times$0.5\arcsec&N/A\\
\hline
\end{tabular}
\tablenotetext{$\dag$}{The peak temperature of IRIS SJ passbands are estimated to be the formation temperature of the major emission lines in the passbands.}
\tablenotetext{*}{The cadence of the HMI magnetograms earlier than 18:00\,UT is 90\,s in order to reduce the load of data for downloading.}
\end{table}

\begin{table}[!ht]
\centering
\caption{ Spectral lines analysed in the present study as taken by IRIS and Hinode/EIS.}
\label{tab_dsp}
\begin{tabular}{c c c c c c}
\hline
Inst. & ion & wavelength &$T_{max}$ (K)& step size & slit pixel\\
\hline
\hline
IRIS &Mg\,{\sc ii}&2796.4\,\AA&$1.4\times10^4$&0.35\arcsec&0.17\arcsec\\
&Mg\,{\sc ii}&2803.5\,\AA&---&---&---\\
  &C\,{\sc ii}&1334.5\,\AA&$2.5\times10^4$&---&---\\
  &C\,{\sc ii}&1335.7\,\AA&---&---&---\\
  &Si\,{\sc iv}&1393.8\,\AA&$7.9\times10^4$&---&---\\
  &Si\,{\sc iv}&1402.8\,\AA&---&---&---\\
  &O\,{\sc iv}&1399.8\,\AA&$1.4\times10^5$&---&---\\
  &O\,{\sc iv}&1401.2\,\AA&---&---&---\\
\hline
EIS &He\,{\sc ii}&256.3\,\AA$^b$&$5\times10^4$&2\arcsec&1\arcsec\\
&Fe\,{\sc viii}$^{*}$&186.6\,\AA&$4.5\times10^5$&---&---\\
&Fe\,{\sc ix}$^{*}$&197.8\,\AA&$7.1\times10^5$&---&---\\
&Fe\,{\sc x}$^{*}$&184.5\,\AA&$1.1\times10^6$&---&---\\
&Fe\,{\sc xi}$^{*}$&180.4\,\AA&$1.4\times10^6$&---&---\\
&Fe\,{\sc xii}&186.9\,\AA&$1.6\times10^6$&---&---\\
&Fe\,{\sc xii}$^{*}$&195.1\,\AA&---&---&---\\
&Fe\,{\sc xiii}$^{*}$&202.0\,\AA&$1.8\times10^6$&---&---\\
&Fe\,{\sc xiii}&203.8\,\AA$^b$&---&---&---\\
&Fe\,{\sc xiv}$^{*}$&270.5\,\AA$^b$&$2.0\times10^6$&---&---\\
&Fe\,{\sc xv}$^{*}$&284.2\,\AA$^b$&$2.2\times10^6$&---&---\\
&Fe\,{\sc xvi}$^{*}$&263.0\,\AA&$2.8\times10^6$&---&---\\
\hline
\end{tabular}
\tablenotetext{}{IRIS spectrometer was observing with an exposure time of 15\,s, and EIS was observing with an exposure time of 45\,s. 
The width of the IRIS slit is 0.35\arcsec\ and that of the EIS slit is 2\arcsec.
The spectral lines blended with lines from different species of ions are denoted by $^b$.
The EIS spectral lines denoted by asterisks are used in DEM analysis.}
\end{table}

\par
IRIS was running in a 320-step dense raster mode, and the spectrograph obtained spectral data of a region with a size of 112\arcsec$\times$129\arcsec\  and a center at (x=--3.3\arcsec, y=33.2\arcsec).
Hinode/EIS was scanning a region centering at (x=--21.2\arcsec, y=--26.6)  in short wavelength (SW) band with a size of 120\arcsec$\times$512\arcsec.
The Hinode/XRT field-of-view has a size of 384\arcsec$\times$384\arcsec\ and centers at (x=--3.0\arcsec, y=--0.5).
A summary of the imaging data analysed in this study is given in Table\,\ref{tab_dimg}, and
the information of spectral data is shown in Table\,\ref{tab_dsp}.

\par
All the data were first prepared with standard procedures provided by the instrument teams. The coalignment of the data were obtained by cross-correlations of observations at wavelength with closest representative temperatures.
In the present case, the referent coordinates are given by HMI magnetograms. The AIA 1700\,\AA\ images were used to align with the HMI magnetograms, and the rest AIA passbands were aligned to each other via cross-correlations.  
The IRIS 1330\,\AA\ data were then aligned to AIA 1700\,\AA\ images.
The XRT Al-poly images were aligned to AIA 94\,\AA, and
Hinode EIS Fe\,{\sc xii}\,195\,\AA\ raster images are then aligned to soft X-ray.

\par
\begin{figure*}
\includegraphics[clip,trim=1cm 0cm 0cm 0cm,width=\textwidth]{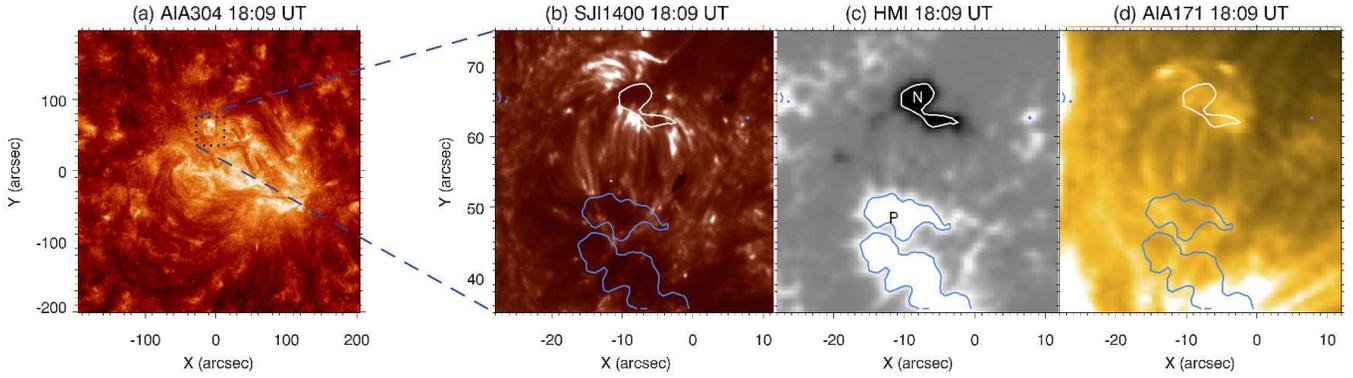}
\caption{(a): an area near the center of the solar disc viewed in AIA 304\,\AA\ passband at 18:09\,UT on 2017 August 26, in which the region of interest with cool transition region loops studied in the present work is enclosed by dotted lines.
(b)--(d): The IRIS SJ 1400\,\AA\ image (b), the HMI magnetogram (c) and the AIA 171\,\AA\ image (d) of the region of interest,
on which the white and blue contour curves represent magnetic flux density measured at 18:09\,UT with levels of $-800$\,Mx\,cm$^{-2}$ and $800$\,Mx\,cm$^{-2}$, respectively.
The HMI magnetogram is artificially saturated at $-1000$\,Mx\,cm$^{-2}$ (black) and $1000$\,Mx\,cm$^{-2}$ (white).
The positive and negative polarities that are connected by the magnetic loops are marked in the magnetogram as ``P'' and ``N'', respectively.}
\label{fig:fov}
\end{figure*}

The region of interest where a group of magnetic loops were observed and its context on AIA 304\,\AA\ passband are shown in Figure\,\ref{fig:fov}.
The region is part of the active region NOAA AR12672.
By the beginning of the observations, the active region itself had been fully grown and no large-scale magnetic flux emergence was observed in this period of time.
In the region of interest, small-scale flux emergence in the later phase was observed in the current dataset (see details in Section\,\ref{subsect_mag}). This small flux emergence formed formed a small loop system that is the main object of the present work.
This is different from many previous studies on flux emergences, which focused on emerging stage of the active regions themselves.

\begin{figure}
\includegraphics[clip,trim=0.5cm 0cm 0cm 0cm,width=\linewidth]{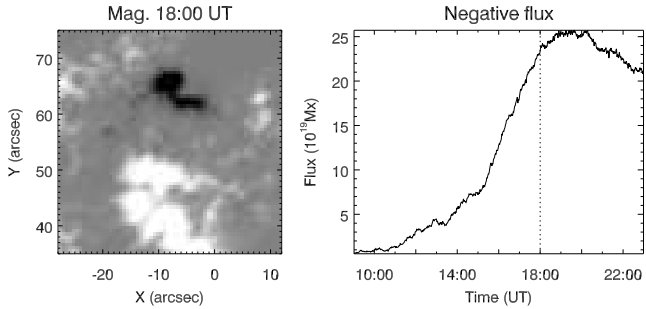}
\caption{Evolution of HMI magnetograms.
Left: HMI magnetogram of the region, which is scaled from $-1000$\,Mx\,cm$^{-2}$ (black) to $1000$\,Mx\,cm$^{-2}$ (white). 
Right: the variation of the total magnetic flux of negative polarity in the region as shown on the left.
The dotted line on the right panel indicates the observing time of the magnetogram shown in the left panel. (An animation is provided online.)}
\label{fig:maglc}
\end{figure}

\subsection{Analysis of spectroscopic data}
IRIS data used in the present study are level 2 data, which have been fully prepared by the instrument team and no further calibration is applied.
EIS data were downloaded as level 0.
They were firstly reduced by the standard procedures packaged in {\it eis\_prep.pro},
which also includes radiometric calibration.
The details of the procedures could be found in the document of {\it eis\_swnote\_01} in the {\it solarsoft} directories.

\par
To derive line parameters of the observed spectral lines, including peak intensity, line centre and line width, Gaussian fits are used.
For those spectral lines used to derive Doppler velocities, nonthermal velocities and electron density, blending lines should be carefully handled.
Most of these lines could be well fitted by single Gaussian function even though some are blended.
In these cases, the blended lines could not be removed with multi-Gaussian fits and their effects will be considered while understanding the observational results.
The only case that the blending line could be removed by double Gaussian fits is Fe\,{\sc xiii}\,203.8\,\AA\ that is blended by Fe\,{\sc xii}\,203.7\,\AA.

\par
To obtain Doppler velocities, wavelength calibration is required.
The wavelength calibration has been performed in IRIS level 2 data by the instrument team using reference of neutral lines (see details in IRIS Technical Note 20).
However, absolute wavelength calibrations for the EIS data are difficult.
Alternatively, an averaging profile from a quiet-sun region could be used as reference\,\citep{2012ApJ...744...14Y}.
However, this quiet-sun method cannot be used in the present because very few quiet-sun region is covered in the field-of-view.
In the present study, a spectral profile averaging of the entire field-of-view was first obtained and its line center was taken as the rest wavelength of that spectral line.
In this case, the obtained Doppler velocities from EIS data should have an accuracy of about 4.4\,\kms\,\citep{2010SoPh..266..209K}. 
Comparing to calibration with quiet Sun region,
this wavelength calibration method could lead to an offset of 5.4\,\kms\ toward red-shifts in the corona\,\citep[see discussion in][]{2012ApJ...744...14Y},
and this offset has been taken into account in the present study.

\par
While constructing a map for Doppler velocities, thermal drifts and slit tilt should also be corrected.
For EIS data, the thermal drifts could be obtained through a method proposed by \citet{2010SoPh..266..209K} and the slit tilt could be obtained from specially-designed observations.
These corrections could be achieved by softwares provided in the {\it solarsoft} by the instrument team and they have been included for producing the Doppler velocity maps.
For IRIS data, the basic corrections for thermal drifts and slit tilts have been applied by the data processing pipeline and no further corrections are required in the present case. 

\par
To derive nonthermal velocities, the calculation steps given in \citet{1998ApJ...505..957C} are followed.
The instrument broadening given by the instrument team are used, which are  0.054\,\AA\ (FWHM) for EIS Fe\,{\sc xii} 195.1\,\AA\ and Fe\,{\sc xiii} 202.0\,\AA\ and 0.032\,\AA\ (FWHM) for IRIS \siiv\ 1393.8\,\AA.
The thermal temperatures of a spectral line is assumed to be the formation temperatures as listed in Table\,\ref{tab_dsp}.

\section{Results and discussion}
\label{sect_results}
In this section, I show the observational analysis of the magnetic loops using multi-wavelength imaging and spectroscopic data.

\subsection{Evolution of the magnetic features}
\label{subsect_mag}
In Figure\,\ref{fig:fov}, one can see that the magnetic loops are rooted in a large patch of positive polarity in the south (marked as ``P'' in Figure\,\ref{fig:fov}c) and a patch of negative polarity in the north (marked as ``N'' in Figure\,\ref{fig:fov}c).
The positive polarity is corresponding to a set of small sunspots that were part of the following sunspot group of the active region.
The evolution of magnetic features in the region are shown in Figure\,\ref{fig:maglc} and the associated animation.
We can see that the negative polarity clearly shows an emerging process from 10:00\,UT to 20:00\,UT.
The total negative flux in the region increases about 25 times from about $10^{19}$\,Mx,
which gives an flux-emerging rate of $\sim7\times10^{15}$\,Mx\,s$^{-1}$ in average.
During the emerging process, small magnetic features with negative polarity were appearing at the edge of the major positive polarity (``P'' in Figure\,\ref{fig:fov}c)
and then moved northward and merged each other to form the large patch as seen in Figure\,\ref{fig:fov}c (denoted as ``N'').
Since the positive polarity is predominant in the region, its emergence was not as clearly as that of the negative one in the observations.

\par
\begin{figure*}
\includegraphics[clip,trim=1.5cm 3cm 0cm 0cm,width=\linewidth]{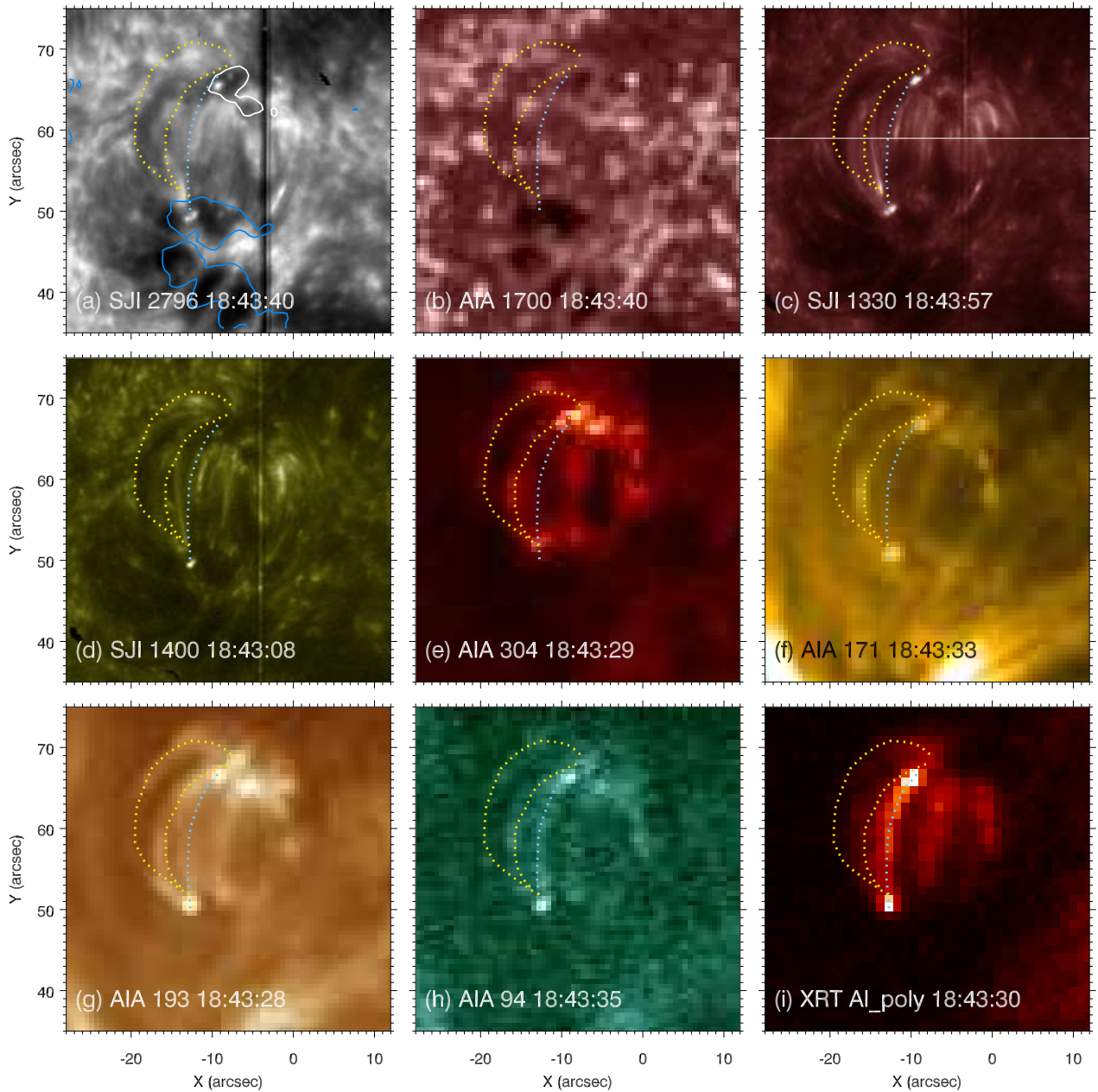}
\caption{The loop system viewed in multiple passbands of IRIS, AIA and XRT.
The blue and white solid lines in panel (a) are the contours of magnetic flux densities measured at 18:43\,UT with levels of $800$\,Mx\,cm$^{-2}$ and $-800$\,Mx\,cm$^{-2}$, respectively.
The cyan dotted lines denote a loop thread that is clearly seen in XRT image, and the yellow dotted lines denote two loop threads that are clearly seen in AIA 304\,\AA\ image.
The white solid line in panel (c) indicates the cut, where the variations of radiation in the observing passbands are shown in Figure\,\ref{fig:clplc}.
(An animation is provided online.)}
\label{fig:mulimgs}
\end{figure*}

As seen in the variation of the negative flux (right panel of Figure\,\ref{fig:maglc}), the IRIS and Hinode observations were taken at the late phase of the emergence, 
when many magnetic loops have already formed between the two major polarities.
During this period of observing time (i.e. 18:00\,UT -- 19:27\,UT), many small negative polarities were still appearing at the edge of the major positive polarity and moving toward the major negative one.
Meanwhile, we could also see many small positive polarities moving into the region between the positive and negative polarities.
These small positive polarities might have appeared at the edge of the major negative polarity or split from the major positive polarity.
Most of these small polarities (both positive and negative) could not reach the major ones.
They might simply disappear on the half way; might also meet and cancel each other.
These emerging, splitting and cancelling of small polarities could be well traced by the small dynamic brightenings in AIA 1700\,\AA\ images (see the animation associated with Figure\,\ref{fig:mulimgs}).

\begin{figure}
\includegraphics[clip,trim=0.5cm 4cm 0cm 0cm,width=\linewidth]{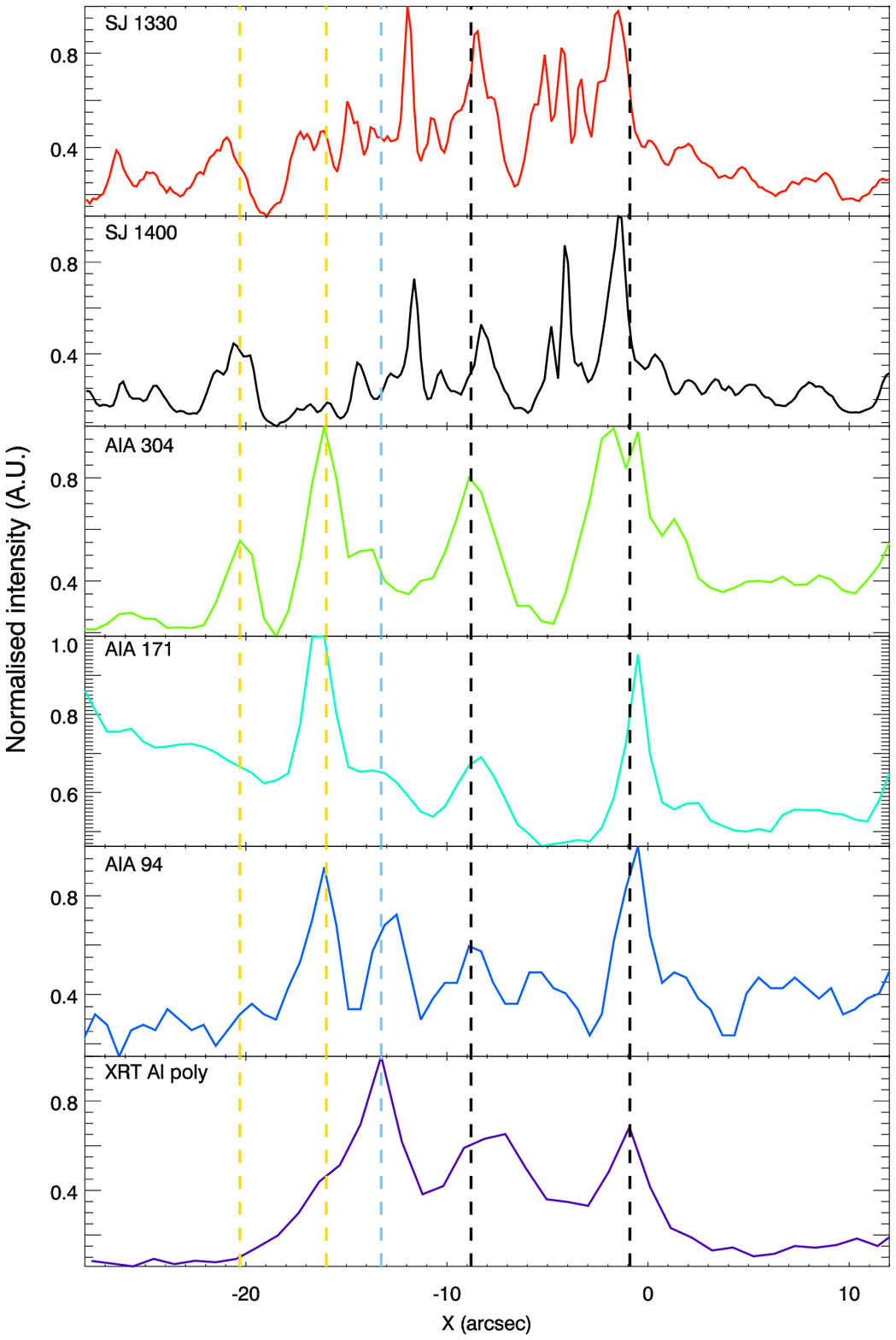}
\caption{The variations of radiation in multiple passbands along the cut marked in Figure\,\ref{fig:mulimgs}c. The yellow dashed lines indicate the locations of the loop threads that are marked as yellow dotted lines in Figure\,\ref{fig:mulimgs}. The cyan dashed lines indicates the location of the loop thread that are marked as cyan dotted lines in Figure\,\ref{fig:mulimgs}. The black dashed lines denote locations of another two loop threads that could be identified by all the AIA EUV channels and XRT channel.}
\label{fig:clplc}
\end{figure}

\subsection{Imaging of the magnetic loop system}
\label{subsect:imaging}
In Figure\,\ref{fig:mulimgs} and the associated animation, I show the evolution of the magnetic loop system viewed in multiple passbands of the imagers.
The details of the loop system could be well seen in the IRIS SJ 1330\,\AA\ and 1400\,\AA\ images.
While many loops were connecting the two major polarities, we can also see that some others were connecting the major polarities and the weaker ones.
The loop system was very dynamic and evolved (appearing and disappearing) very fast in time scale comparable to the cadences of the observations.
Brightenings frequently occurred in the footpoint regions and led to flaring of many loop threads.
However, there are also some flaring loop threads do not show bright footpoints in the IRIS SJ 1330\,\AA\ and 1400\,\AA\ images.
In general, the loop system shows clear response in images of IRIS SJ passbands, AIA EUV channels and XRT Al\_poly filter.
However, the structures viewed in AIA and XRT are much more fuzzy than that in IRIS SJ passbands.
One of the possible reasons is the lower spatial resolutions of the AIA and XRT images,
and another reason could be that the counterparts of nearby loop threads at higher temperatures cannot be separated from each other.

\par
The higher resolution images from IRIS SJ 1330\,\AA\ and 1400\,\AA\ (Figures\,\ref{fig:mulimgs}c\&d) show many fine threads of loops with a cross-section $\lesssim0.5$\arcsec\ (full width at half maximum) and $\sim$12\arcsec\ separation between the two footpoints.
Beside these loops connecting the two major polarities, we also see smaller ones with length of about 5\arcsec\ that are connecting a major polarity with minor ones in between (see the animation associated with Figure\,\ref{fig:mulimgs}).
For comparison, the cross-sections of the loop threads determined from the AIA 171\,\AA\ and XRT images are around 2\arcsec (Figures\,\ref{fig:mulimgs}\&\ref{fig:clplc}).
The width of the hot threads are consistent with those measured in TRACE filter images\,\citep{2005ApJ...633..499A}. Such hot threads seen in AIA data might contain multiple finer threads as reported in Hi-C data\,\citep{2013ApJ...772L..19B}, and this could be the case in the present loop system although we do not have higher-resolution images at hot temperatures.

\par
We also notice that not all loop threads in lower temperatures have visible counterparts in higher temperatures.
In Figure\,\ref{fig:clplc}, I show light curves taken along a slit across the loop system seen at around 18:43:30\,UT (shown as the white solid line in Figure\,\ref{fig:mulimgs}c).
A peak in these curves is representative of a loop thread seen in the corresponding channel.
Many loop threads seen in the IRIS SJ 1330\,\AA\ and 1400\,\AA\ passbands do not have distinguishable counterparts in the images taken in passbands with hotter representative temperatures (see Figures\,\ref{fig:mulimgs}\&\ref{fig:clplc}), 
which is consistent with previous studies\,\citep[e.g.][]{1997SoPh..175..487F,2000A&A...359..716S}.
In Figure\,\ref{fig:mulimgs}, I highlight a loop thread (outlined by the cyan dotted lines) that is clearly seen in X-ray and AIA 94\,\AA\ images but almost invisible in the other passbands.
(This could also be seen in Figure\,\ref{fig:clplc}, the peaks in the light curves denoted by cyan dashed line).
In contrast, two loop threads (outlined by yellow dotted lines in Figure\,\ref{fig:mulimgs}) are clearly visible in AIA 304\,\AA\ and 171\,\AA\ but almost invisible in X-ray.
(This could also be seen in Figure\,\ref{fig:clplc}, the peaks in the light curves denoted by yellow dashed lines).
This suggests that these loop threads have counterparts only in a narrow temperature range,
and this is consistent with that reported by \citet{2008ApJ...686L.131W}.

\par
There are also some loop threads having clear counterparts in all the passbands.
We could see that two peaks indicated in Figure\,\ref{fig:clplc} by black dashed lines are corresponding to two loop threads identified on the images of all AIA EUV channels and XRT Al-Poly filter.
By taking into account the spatial resolution of the instruments, we can also consider that these two loop threads also correspond to peaks in IRIS SJ 1330\,\AA\ and 1400\,\AA\ light curves (see Figure\,\ref{fig:clplc}).

\begin{figure*}
\includegraphics[clip,trim=1cm 1cm 0.5cm 0cm,width=\linewidth]{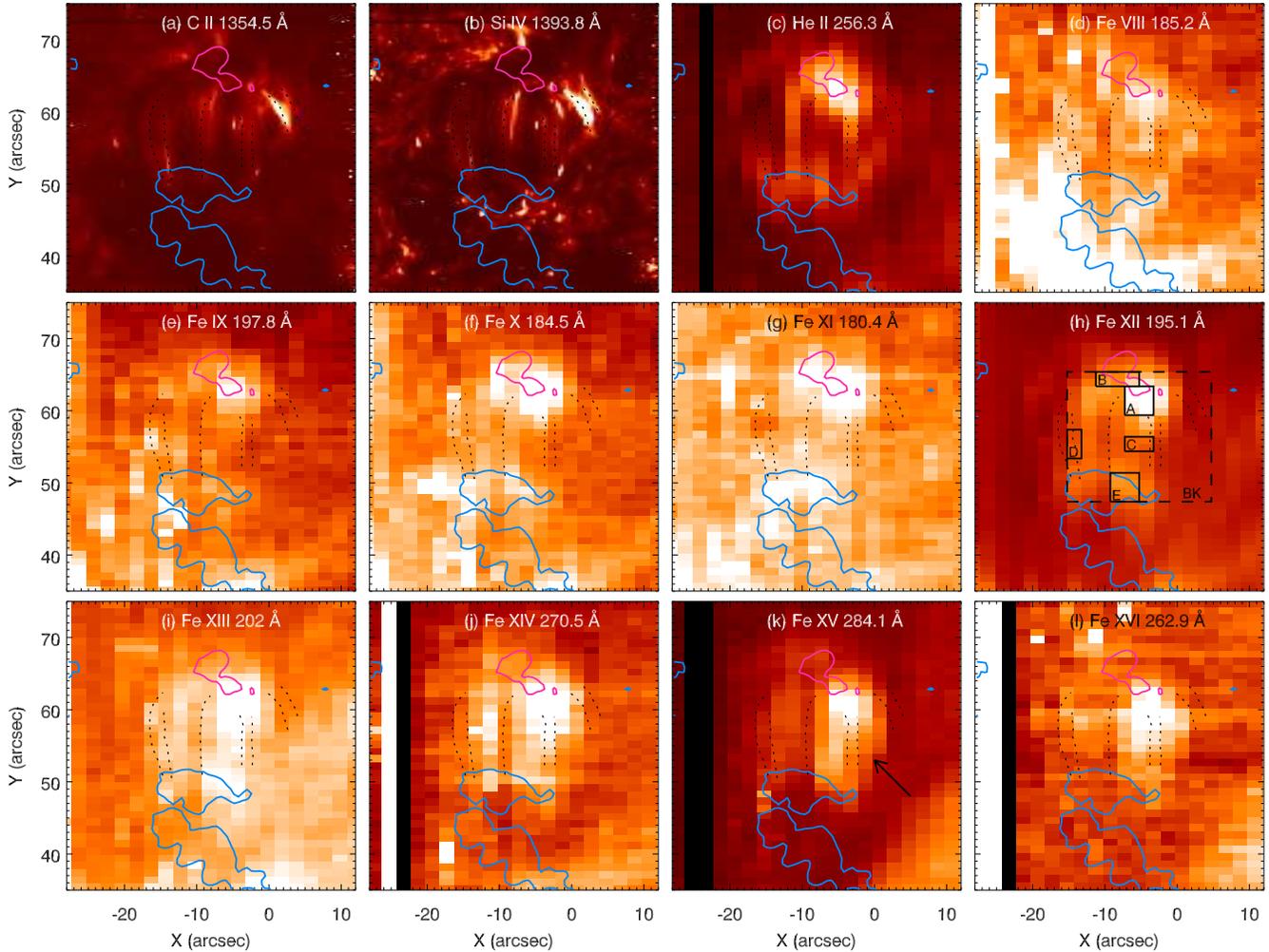}
\caption{The radiance maps of the magnetic loop system taken with IRIS and EIS spectral lines. IRIS scanned this region from 18:25\,UT to 18:56\,UT, and EIS scanned the region in the time period of 19:00--19:17\,UT. 
The blue and purple lines are the contours of magnetic flux densities measured at 18:48\,UT with levels of $800$\,Mx\,cm$^{-2}$ and $-800$\,Mx\,cm$^{-2}$, respectively.
The black dotted lines are tracers of a few magnetic loop threads identified in \siiv\ image (panel b), which could be used as references to compare images taken at different wavelengths.
The radiance from the regions enclosed by the rectangles (solid lines) and denoted by letters of ``A--E'' in panel (h) are used for DEM analysis.
The background/foreground subtraction is obtained from the region enclosed by dashed lines and denoted by ``BK'' (see Section\,\ref{sect_dem} for details).}
\label{fig:sprad}
\end{figure*}

\subsection{IRIS and EIS spectroscopic observations}
This region was scanned by the IRIS slit from 18:25\,UT to 18:56\,UT, and repetitively by EIS slit from 18:00\,UT to 23:59\,UT.
Because IRIS and EIS were scanning the region of interest at different time,
the analysis here focuses on the global characteristics of the loop system rather than any individual loop thread.
In Figure\,\ref{fig:sprad}, I show the IRIS and EIS raster images of the region in a few spectral lines.
With the high-resolution data, IRIS could well distinguish many loop threads in the raster images (Figure\,\ref{fig:sprad}a\&b).
Even though the loop system in EIS data is fuzzy and it cannot distinguish any single loop threads,
we could also see that the region occupied by the loops is brighter than the background (especially in He\,{\sc ii}, Fe\,{\sc xi}--Fe\,{\sc xvi}).
Using the imaging data as guidance, I believe that some of the loop threads (for example, see the fuzzy bright structure denoted by the arrow in Figure\,\ref{fig:sprad}k) have been heated to more than $2\times10^6$\,K (i.e. formation temperatures of Fe\,{\sc xv} and Fe\,{\sc xvi}).
These hot loops are seemingly not visible in the raster images of Fe\,{\sc viii}--Fe\,{\sc x} (Figure\,\ref{fig:sprad}d--f).
It again confirms that most of the loop threads have a narrow range of temperatures.

\par
In \cii\ and \siiv\ radiance maps (Figure\,\ref{fig:sprad}a\&b), the footpoints of the loop system appear to be dark region that suggests lower emission than background (see the blue and purple contours in the figure).
The north footpoint (purple contour) are most clearly seen in He\,{\sc ii}, Fe\,{\sc x} and Fe\,{\sc xi} images (Figure\,\ref{fig:sprad}c,f,g), though its southwest portion is clearly seen in the other EIS spectral lines.
While He\,{\sc ii} 256.3\,\AA\ is blended by Si\,{\sc x} and Fe\,{\sc x}, it suggests that the temperature of the north footpoint is in the range of $1.1\times10^6$--$1.4\times10^6$\,K (i.e. between formation temperatures of Fe\,{\sc x} and Fe\,{\sc xi}).
The south footpoint (blue contour) is most clearly seen in Fe\,{\sc xi} (Figure\,\ref{fig:sprad}g), suggesting its temperature at about $1.4\times10^6$\,K.
This suggest that heating was concentrated in the footpoints of the loop system.
To further understand these loops, I exploit the spectroscopic data to deduce their physical parameters and show in the following.

\begin{figure*}
\includegraphics[clip,trim=0cm 1cm 0cm 0cm,width=\linewidth]{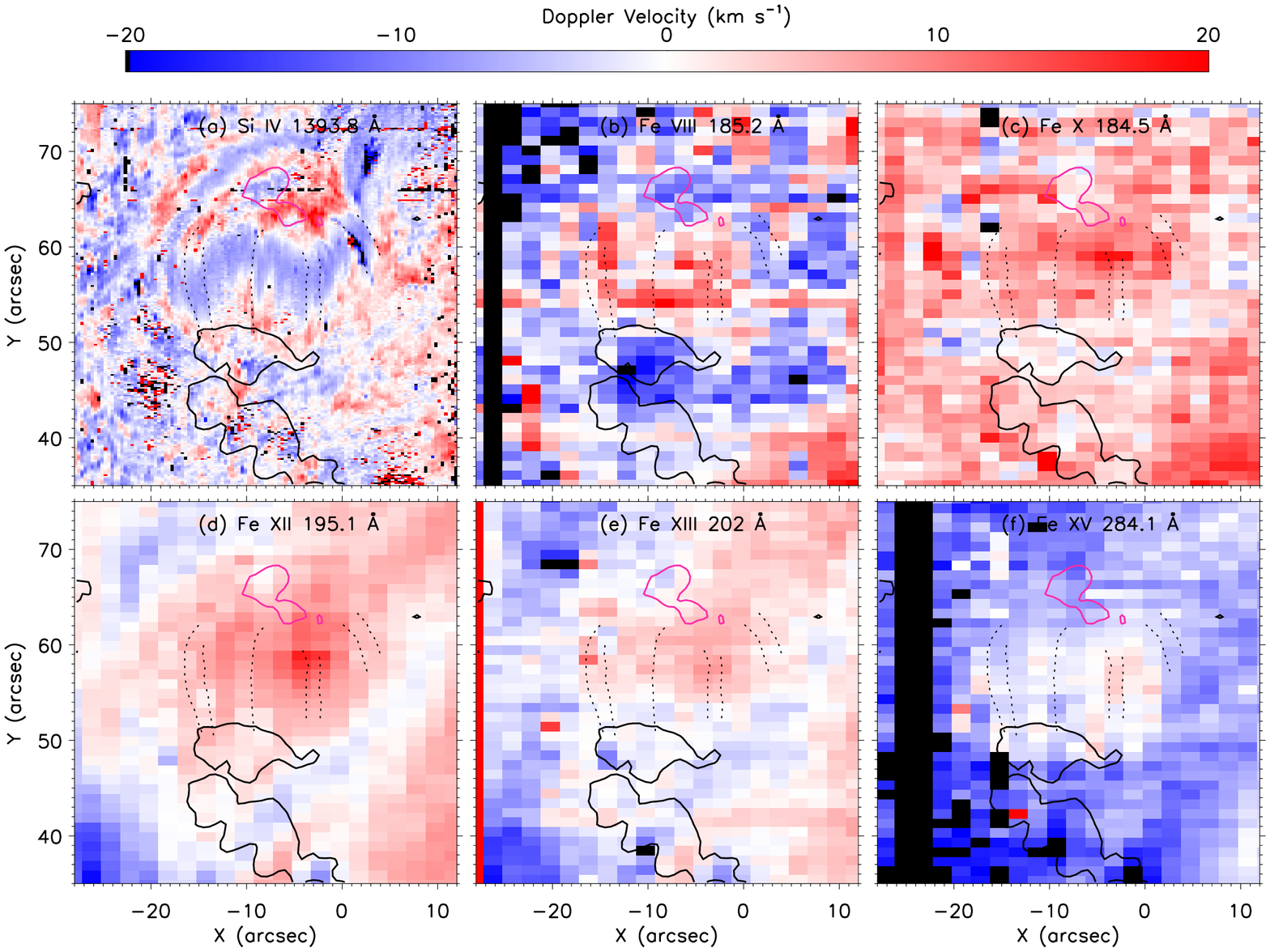}
\caption{Dopplergrams of the magnetic loop system measured with IRIS \siiv\,1393.8\,\AA\ and EIS Fe\,{\sc viii}\,185.2\,\AA, Fe\,{\sc x}\,184.5\,\AA, Fe\,{\sc xii}\,195.1\,\AA, Fe\,{\sc xiii}\,202.0\,\AA\ and Fe\,{\sc xv}\,284.1\,\AA. 
The observing time was as same as that given in Figure\,\ref{fig:sprad}. 
The purple contour lines indicate the magnetic flux density at  $-800$\,Mx\,cm$^{-2}$ and the black ones are representative of that at $800$\,Mx\,cm$^{-2}$. 
The black dotted lines are tracing a few loop threads identified in the \siiv\ radiance image as shown in Figure\,\ref{fig:sprad},
which could be used as references to compare different images.}
\label{fig:spdopp}
\end{figure*}

\begin{figure*}
\includegraphics[clip,trim=0cm 0cm 0cm 0cm,width=\textwidth]{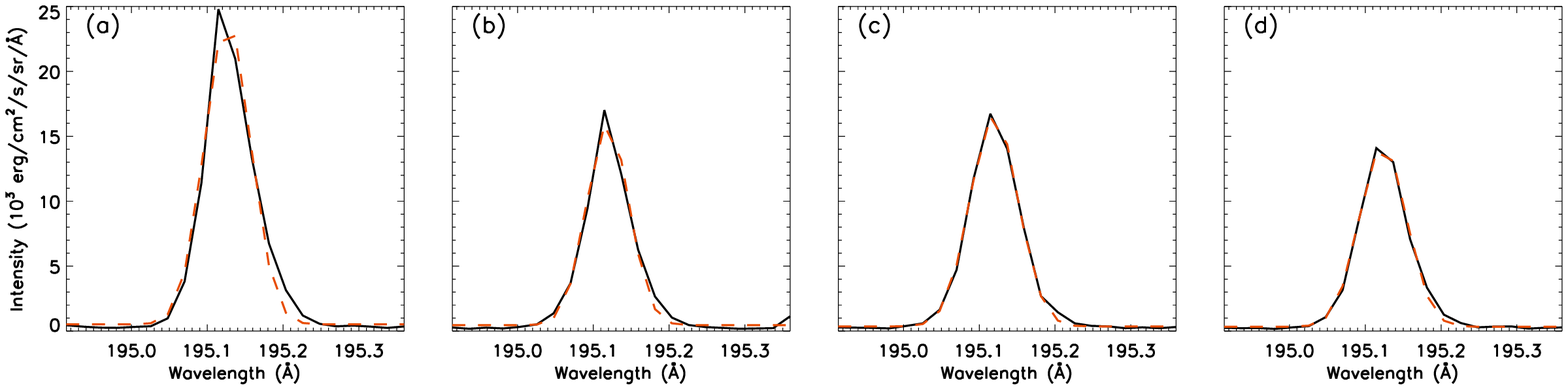}
\caption{A few samples of Fe\,{\sc xii}\,195.12\,\AA\ profiles from the region of interest observed by EIS.
The observed profiles are shown as black solid lines, and the single Gaussian fits to the observed profiles are shown as red dashed lines.
}
\label{fig:fe12samp}
\end{figure*}

\subsubsection{Doppler velocities}
To derive the Doppler velocities, a spectral profile averaging of the entire field-of-view was first obtained and its line center was taken as the rest wavelength of that spectral line.
In Figure\,\ref{fig:spdopp}, I display the Doppler maps of the region measured with different spectral lines by IRIS and EIS.
In IRIS \siiv\ 1393.8\,\AA\ (transition region temperature), the loop system shows systematic blue-shifted ($\sim$10\,\kms) in the middle of the loops, while their footpoints are red-shifted (Figure\,\ref{fig:spdopp}a).
The blue-shifts in the middle of the loops are in line with the picture of flux emergence.
Because the blue-shifted pattern is systematic in the loop region, it indicates that the loop system should keep emerging during the period of the IRIS raster ($\sim$25\,minutes).
This should not be the case for any particular loop thread, otherwise it would move upward for about 20\arcsec\ and this should be observed in the imaging data (unless they are all moving along line-of-sight).
The systematic blue-shifts could be understood if there are many loops continuously moving upward in the transition region.
Because no clear upward motion was observed in the imaging data, these loops should not moving very far while they are having transition region temperatures.
The absence of downflows (red-shifts) could be due to heating in the loops that leads to disappearance of the loops in the transition region temperature.
This understanding is supported by the extremely dynamic nature (frequently appearring and disappearring) of the loops seen in the IRIS SJ 1330\,\AA\ and 1400\,\AA\ images (see section\,\ref{subsect:imaging}).
Because these loops have relatively small range of temperatures, the continuing emergence could be observed as frequent appearance of the loop threads in these passbands and heating of the loop threads could lead to their disappearance.
Therefore, the downflows could only be seen in higher temperatures,
and these are observed in EIS Fe\,{\sc viii}--Fe\,{\sc xv} (Figure\,\ref{fig:spdopp}b--f).

\par
With EIS data, Doppler maps are derived from Fe\,{\sc viii} 185.2\,\AA\ (Figure\,\ref{fig:spdopp}b), Fe\,{\sc x} 184.5\,\AA\ (Figure\,\ref{fig:spdopp}c), Fe\,{\sc xii} 195.1\,\AA\ (Figure\,\ref{fig:spdopp}d), Fe\,{\sc xiii} 202.0\,\AA\ (Figure\,\ref{fig:spdopp}e) and Fe\,{\sc xv} 284.1\,\AA\ (Figure\,\ref{fig:spdopp}f), with temperatures ranging from $4.5\times10^5$\,K to $2.2\times10^6$\,K.
In the Doppler maps of Fe\,{\sc viii} to Fe\,{\sc xiii}, it is clear that red-shifts are predominant in the middle of the loops.
The Doppler map of Fe\,{\sc xv} (Figure\,\ref{fig:spdopp}f) also shows clear red-shifted pattern at places around (X=$-$3\arcsec, Y=55\arcsec) where loop structure appears (see Figure\,\ref{fig:sprad}k).
From Fe\,{\sc x} to Fe\,{\sc xv}, the Doppler velocities in the middle of the loops show a trend from large ($\gtrapprox$10\,\kms) to small ($<$5\,\kms\ that is closed to the accuracy of the measurement).
This might imply that the loop plasmas become more stationary while they are heated to higher temperatures.

\par
Note that the Doppler shifts measured with Fe\,{\sc xii}\,195.1\,\AA\ are significantly larger than that with Fe\,{\sc xii} 202.0\,\AA, and this bias might be brought in by the blend of Fe\,{\sc xii} 195.2\,\AA\ at the red-wing of the Fe\,{\sc xii}\,195.1\,\AA\ lines.
While randomly checking tens of locations selected from the region, in agreement with \citet{2009A&A...495..587Y}, I found that the contribution of Fe\,{\sc xii} 195.2\,\AA\ is more significant in the brighter region than that in average (see a few examples in Figure\,\ref{fig:fe12samp}).
However, the blending cannot be easily removed in most of the positions of the region because their profiles are well fitted by single Gaussian functions (see examples in Figures\,\ref{fig:fe12samp}c\&d).
Although the blending might be removed by double Gaussian fits with assumptions of a few parameters of the blending component\,\citep[see e.g.][]{2016ApJ...827...99T}, it could easily bring in additional artificial effect and also the fittings cannot converge in many cases.
Nevertheless, the Fe\,{\sc xii}\,195.1\,\AA\ Doppler map still gives a good indication of velocity at the corresponding temperature while we have the Fe\,{\sc xii} 202.0\,\AA\ one for reference.

\begin{figure}
\includegraphics[clip,trim=0cm 0cm 0cm 0cm,width=\linewidth]{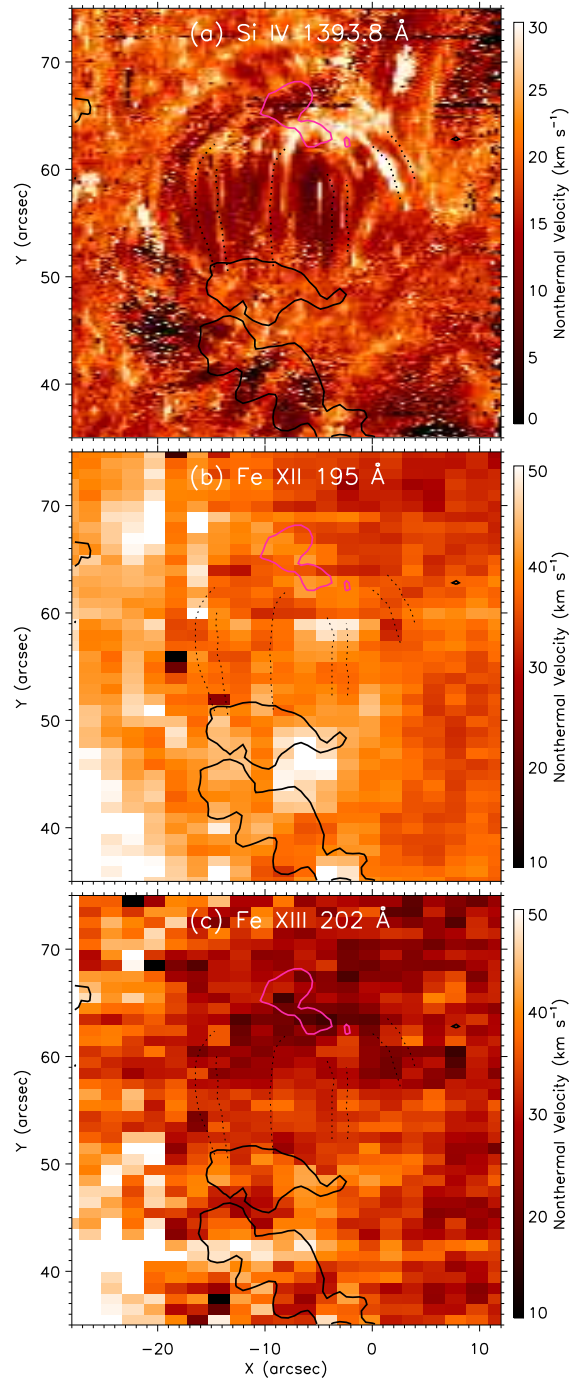}
\caption{Nonthermal velocities of the region measured with IRIS \siiv\ 1393.8\,\AA\ (panel a), EIS Fe\,{\sc xii} (panel b) and Fe\,{\sc xiii} (panel c). The purple and black contour lines outline the magnetic polarities, and the black dotted lines are tracing a few loop threads identified in the \siiv\ radiance image as shown in Figure\,\ref{fig:sprad}.}
\label{fig:spwid}
\end{figure}
\subsubsection{Nonthermal velocities}
We measured the nonthermal velocities of the region by assuming the thermal temperature to be the formation temperatures (listed in Table\,\ref{tab_dsp}).
In Figure\,\ref{fig:spwid}, I display the nonthermal velocities of the region measured with IRIS \siiv\,1393.8\,\AA\ and EIS Fe\,{\sc xii}\,195.1\,\AA\ and Fe\,{\sc xiii}\,202.0\,\AA.

\par
In IRIS \siiv\,1393.8\,\AA, a few locations where flaring loops exist have nonthermal velocities of $\sim$30\kms, which is significantly larger than that of the surrounding background.
Apart from those locations, the nonthermal velocities of the loop region are less than 10\,\kms, much smaller than that of the surrounding region.
This suggests that most of the emerged loop plasma was less disturbed by nonthermal processes in the transition region before flaring-up.
The small nonthermal velocity is dominant in the loops, suggesting that most of the emerging loops have not been heated (by nonthermal processes) before reaching the transition region.

\par
Using EIS data, nonthermal velocities from Fe\,{\sc xii}\,195.1\,\AA\ and Fe\,{\sc xiii} 202.0\,\AA\ are derived (Figure\,\ref{fig:spwid}b\&c).
In the loop region, the nonthermal velocities derived from Fe\,{\sc xii}\,195.1\,\AA\ are 
in the range of 40--50\,\kms, which are overall larger than that from Fe\,{\sc xiii} 202.0\,\AA\ (30--40\,\kms).
This discrenpancy is again due to the blending of Fe\,{\sc xii} 195.2\,\AA\ at the red wing of Fe\,{\sc xii}\,195.1\,\AA.
Please note that radiation calibration has been applied on the present data, which might overestimate the values by 10--20\%\,\citep{2016ApJ...820...63B,2016ApJ...827...99T}.
Taking into account this effect, the nonthermal velocities measured here are still larger than that in normal active regions\,\citep[$\sim20$\,\kms, see e.g.][]{2016ApJ...820...63B,2016ApJ...827...99T}.
Since these loops are active and have similar size as coronal bright points, the nonthermal velocities of these loops in corona are consistent with that of active region bright points\,\citep{2009ApJ...705L.208I}.

\par
Nevertheless, both nonthermal velocity maps from EIS data indicate that the entire loop region has similar nonthermal velocities as the surrounding quiet corona.
From the point of nonthermal velocity, this indicates that the emerged loop plasma is not much different from the normal coronal plasma.
If the emerging loops have been heated by any nonthermal processes, they should have occurred before reaching these temperatures.

\subsubsection{Electron density measured with coronal lines}
\begin{figure*}
\includegraphics[clip,trim=0cm 0cm 0cm 0cm,width=\linewidth]{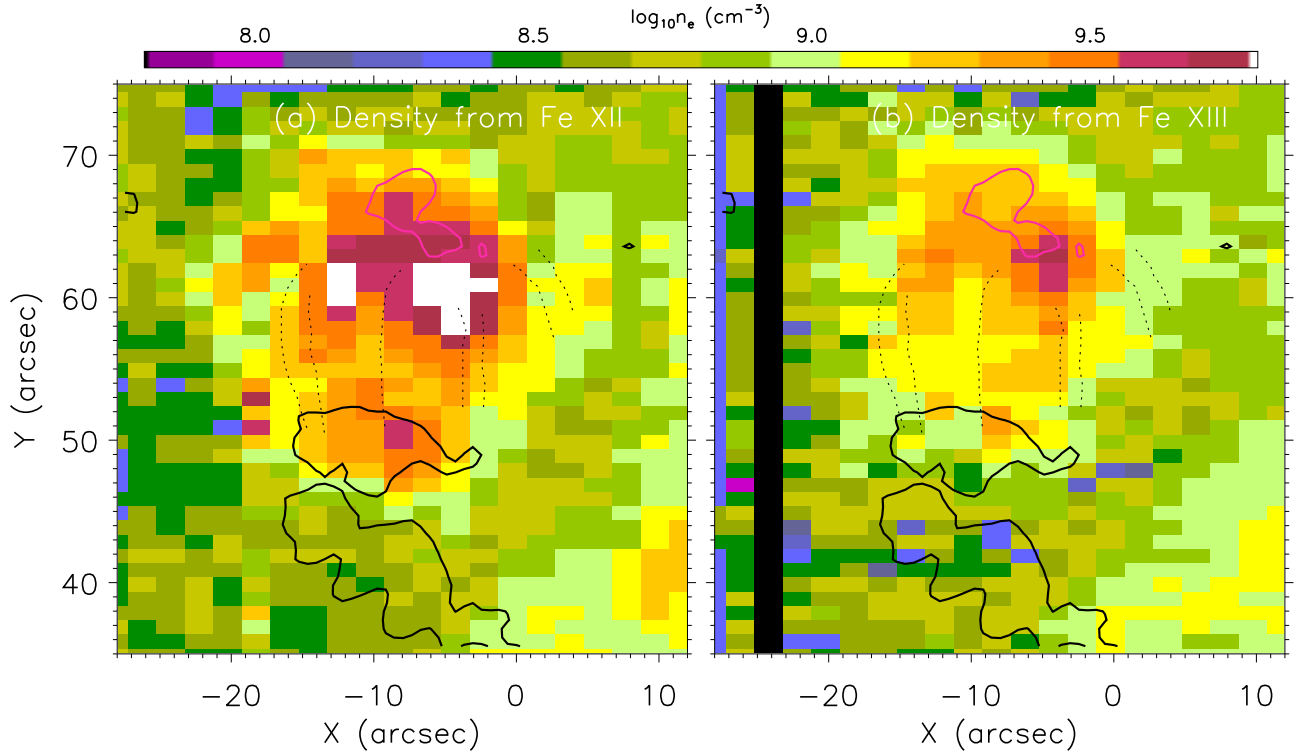}
\caption{Electron-density maps of the loop region measured with EIS line pairs of Fe\,{\sc xii} $\lambda\lambda$186.9/195.1 and Fe\,{\sc xiii} $\lambda\lambda$ 202.0/203.8.
The contours and the dotted lines are as same as those shown in Figures\,\ref{fig:sprad}--\ref{fig:spwid}.}
\label{fig:eisne}
\end{figure*}

Using EIS spectroscopic data, I produce electron-density maps (Figure\,\ref{fig:eisne}) of the loop region using the intensity ratios of the line pairs of Fe\,{\sc xii} $\lambda\lambda$186.9/195.1 and Fe\,{\sc xiii} $\lambda\lambda$ 202.0/203.8\,\citep[both are recommended for density diagnostics, see][]{2007PASJ...59S.857Y,2009A&A...495..587Y} and version 8.0.7 of the CHIANTI atomic database\,\citep{1997A&AS..125..149D,2015A&A...582A..56D}.
We could see that electron density derived from Fe\,{\sc xii} $\lambda\lambda$186.9/195.1 is larger than that from Fe\,{\sc xiii} $\lambda\lambda$ 202.0/203.8, by a factor of $\sim$2.
This discrepancy between the two line pairs has been discussed in detail by \citet{2009A&A...495..587Y}.
The discrepancy should be much less since the new model used in this version of CHIANTI database\,\citep{2015A&A...582A..56D}.
However, this discrepancy has been found to be worse in the EIS data obtained in recent years, and cannot be corrected by the new atomic database alone (Giulio Del Zanna, private communications). 
The discrepancy appears to be a result of a combination of many factors that have yet to be fully understood.
The electron density in the loop region is in the range of $2\sim8\times10^9$\,cm$^{-3}$ measured with Fe\,{\sc xii} $\lambda\lambda$186.9/195.1, and in the range of $1\sim4\times10^9$\,cm$^{-3}$ with Fe\,{\sc xiii} $\lambda\lambda$ 202.0/203.8.
These values are significant larger than that in the background corona (where the density is about $0.6\times10^9$\,cm$^{-3}$).
This indicates that these loops have not only been heated to coronal temperature, but also been filled in with denser plasma.
Furthermore, the electron density of the north footpoint is larger than that of the south, by a factor of $\sim$2.

\subsubsection{Electron density along a cool loop}
\begin{figure*}
\includegraphics[clip,trim=0cm 0cm 0cm 0cm,width=\linewidth]{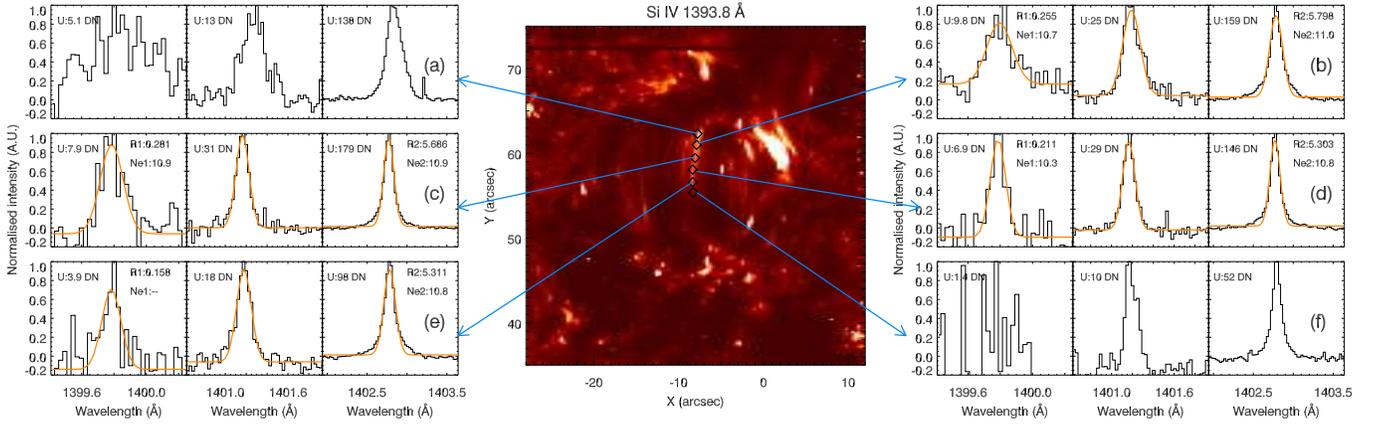}
\caption{The electron density derived in a few regions along a bright transition region loop using IRIS spectroscopic data.
For each region, the O\,{\sc iv} 1399.8\,\AA\ and 1401.2\,\AA\ and \siiv\ 1402.8\,\AA\ (normalised) profiles are displayed. 
The labels following ``U'' give the normalised scale for each profile.
In panels (b--e), the orange lines show the single Gaussian fits to the corresponding profiles.
In panels (b--e), the labels following ``R1'' give the line ratio of O\,{\sc iv} 1399.8\,\AA\ to O\,{\sc iv} 1401.2\,\AA,
and the electron density (in log$_{10}$) derived from this line ratio is shown as the numbers following to the labels of  ``Ne1''.
The labels following to ``R2'' show the line ratio of \siiv\ 1402.8\,\AA\ to O\,{\sc iv} 1401.2\,\AA,
and the electron density (in log$_{10}$) derived from this line ratio is shown as the numbers following to the labels of  ``Ne2''.
These values of electron density are derived with the theoretical model based on the quiet-sun DEM, as shown in Table 2 of \citet{2018ApJ...857....5Y}.}
\label{fig:irisne}
\end{figure*}

In some cases (normally in flaring events), the density sensitive line pair of O\,{\sc iv}\,$\lambda\lambda$1399.8\,\AA/1401.2\,\AA\ are strong enough in IRIS observations and thus could provide an accurate diagnostics of electron density in the solar transition region with unprecedentedly-high resolution\,\citep[see][]{2015arXiv150905011Y,2018ApJ...857....5Y}.
Here, I found that the line pair of O\,{\sc iv} have good signal-to-noise ratio at some locations of a flaring transition region loop (see Figure\,\ref{fig:irisne}).
It gives an opportunity to measure electron density along this transition region loop.
In  Figure\,\ref{fig:irisne}, I show O\,{\sc iv} line profiles of a few locations along the flaring loop.
The length of the loop is about 12\arcsec.
It is apparently flaring at its north portion, and the length of the flaring section is about 8\arcsec.
The flaring portion of the loop shows about 10\,\kms\ blueshift in the Doppler map and about 25\,\kms\ nonthermal velocity in \siiv\,1393.8\,\AA.
In Figure\,\ref{fig:irisne}, we can see that the O\,{\sc iv} 1399.8\,\AA\ line of a few locations in the middle of the loop have relatively good signal-to-noise ratio  (Figure\,\ref{fig:irisne}b--d) and allow density diagnostics using the O\,{\sc iv} line pair.
The electron density measured in these three locations are given as $5.0\times10^{10}$\,cm$^{-3}$ (b), $7.9\times10^{10}$\,cm$^{-3}$ (c) and $1.6\times10^{10}$\,cm$^{-3}$ (d).
These values are consistent with that measured in active region loops by \citet{2016A&A...594A..64P}.

\par
Due to very weak O\,{\sc iv} lines in IRIS data,  the line ratio of  \siiv\ 1402.8\,\AA\ to O\,{\sc iv} 1401.2\,\AA\ are also frequently used for density diagnostics\,\citep[see details in][]{2018ApJ...857....5Y}.
Using the theoretical model based on quiet-sun DEM\,\citep[see Table 2 of][]{2018ApJ...857....5Y}, the electron densities derived from the line ratios of  \siiv\ 1402.8\,\AA\ to O\,{\sc iv} 1401.2\,\AA\ at the three locations are given as $1.0\times10^{11}$\,cm$^{-3}$ (b), $7.9\times10^{10}$\,cm$^{-3}$ (c) and $6.3\times10^{10}$\,cm$^{-3}$ (d).
Even though the line ratios here fall in the range where the line ratios are not much sensitive to the electron densities\,\citep[see Figure 2 of][]{2018ApJ...857....5Y},
I am confident that the electron density measured with the line ratio of  \siiv\ 1402.8\,\AA\ to O\,{\sc iv} 1401.2\,\AA\ is above $5\times10^{10}$\,cm$^{-3}$,
which is in the same magnitude as that from the O\,{\sc iv} line pair.
This suggests that using the line ratios of  \siiv\ 1402.8\,\AA\ to O\,{\sc iv} 1401.2\,\AA\ is an acceptable approach to estimate electron density.
Because the dependence of the line ratios of  \siiv\ 1402.8\,\AA\ to O\,{\sc iv} 1401.2\,\AA\ to electron density could significantly vary in different theoretical models, the loops here are better fitted in the quiet-sun DEM model than the log-linear DEM model\,\citep[see][for details]{2018ApJ...857....5Y}.

\par
The electron density measured with transition region lines is a magnitude larger than that with coronal lines.
It suggests that the loop threads with coronal temperatures is not spatially identical to those with transition region temperatures.
This is in agreement with the imaging data, in which most of the loop threads appear to have narrow range of temperatures.

\subsubsection{Estimation of temperatures}
\label{sect_dem}
\begin{figure}
\centering
\includegraphics[clip,trim=0.2cm 2cm 0.2cm 0.5cm,height=0.9\textheight]{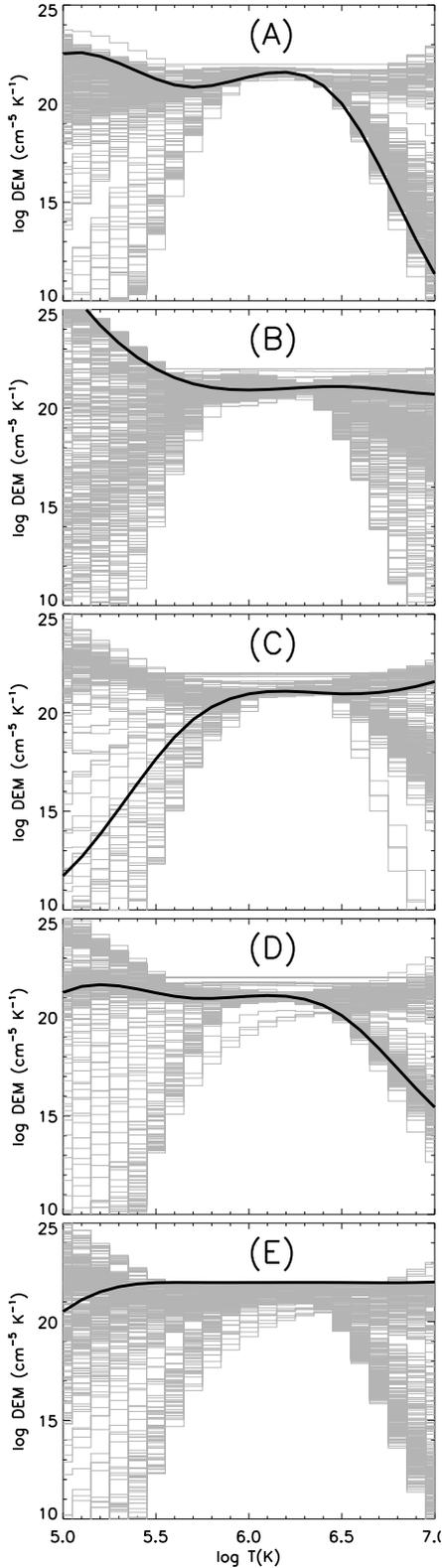}
\caption{The DEM curves for the five regions (``A--E'') marked in Figure\,\ref{fig:sprad}h.
The grey curves are the solutions given by Monte Carlo simulations,
and the best solution is given by the black line.}
\label{fig:eisdem}
\end{figure}

The spectroscopic data provide us with observations in multiple spectral lines at various temperatures.
This gives an opportunity to investigate the temperature profiles of the region using the differential emission measurement (DEM) technique\,\citep[see e.g.][and references therein]{2009ApJ...705.1522B,2011ApJ...730...85B,2012A&A...539A.146H,2015ApJ...807..143C,2018ApJ...856L..17S}.
Here, I used the DEM procedure packed with CHIANTI ({\it chianti\_dem.pro}), in which the {\it XRT\_DEM} package was called to calculate the DEM profiles.
In order to reduce the error brought in by the radiometry calibrations of different instruments, I used the spectral lines from EIS only.
Additionally, I used only the spectral lines from the element of iron to avoid the uncertainty brought in by the abundance.
The spectral lines used for DEM analysis are denoted in Table\,\ref{tab_dsp}, and most of them have been used in previous studies\,\citep[e.g.][]{2009ApJ...705.1522B,2011ApJ...730...85B}.
The spectral lines used in the DEM analysis are formed in temperatures between $4.5\times10^5$--$2.8\times10^6$\,K (i.e. logT/K=5.65--6.45),
therefore, the derived DEM profiles are constrained by the observations only in this temperature range.

\par
Background and foreground subtractions are another important issue that has to be considered in the DEM analysis\,\citep[see e.g.][]{2003A&A...406.1089D,2005ApJ...633..499A}.
A region that is representative of background and foreground is normally selected as close to the analyzed region as possible,
and it should not contain any active features (e.g. other loop threads) in any spectral lines used for DEM analysis.
The background and foreground subtractions become difficult because the region of interest studied here contain many loop threads and they cannot be distinguished in the EIS data.
Here, I performed the DEM analysis on five subregions of the loop system (see Figure\,\ref{fig:sprad}h), including three from the footpoints (``A'', ``B'' and ``E'') and two taken from the middle of the loops (``C'' and ``D'').
I first selected a large region that includes all the analyzed regions (see that denoted as ``BK'' in Figure\,\ref{fig:sprad}h).
While all the pixels in the ``BK'' region are sorted in the order from low to high intensity,
the background and foreground emission is taken by averaging 10\% of all the pixels counting from the lowest intensity one.

\par
The DEM profiles of the analyzed regions are shown in Figure\,\ref{fig:eisdem}.
The Monte Carlo simulations appear to prefer peak temperatures at around logT/K=6.1--6.2 (i.e. $1.3-1.6\times10^6$\,K).
However, we could see that these DEM profiles are not Gaussian with either no clear peak or flat peaks,
therefore, it does not allow quantitively evaluate the temperature distributions\,\citep[see e.g.][]{2008ApJ...686L.131W,2009ApJ...700..762W}.
Such flat peaks in the DEM profiles could be indications of multi-thermal components in the analyzed regions, where loop threads with different temperatures are included.


\subsection{Discussions}
\subsubsection{Flux emergence at its late phase}
While the magnetic flux tube that hosts loop plasma is emerging from the solar convective zone,
it experiences the dramatic change of plasma environment in the solar lower atmosphere,
where the plasma ionisation turns from partial to fully and plasma beta turns from greater than one to less then one.
This might lead to serpentine geometry of flux tube observed in the solar lower atmosphere\,\citep{2001ApJ...554L.111F,2007A&A...467..703C,2008ApJ...687.1373C,2009ApJ...691.1276A,2014LRSP...11....3C,2018ApJ...853L..26H}, 
and this could lead to a variety of energetic events such as Ellerman bombs and UV bursts\,\citep[see e.g.][]{2004ApJ...614.1099P,2009ApJ...701.1911P,2014Sci...346C.315P,2015ApJ...813...86I,2015ApJ...798...19N,2015ApJ...812...11V,2016MNRAS.463.2190N,2016ApJ...824...96T,2017ApJ...836...52Z,2017ApJ...838..101H,2018ApJ...852...95N,2018arXiv180505850Y}.

\par
By analysing observations of a flux emerging region at its early stage, \citet{2017ApJ...836...63T} reported two types of local heating events that could result from magnetic reconnection in the bald-patches (magnetic dips) of the emerging flux and shocks or strong compression caused by fast downflows along overlying magnetic system.
Similarly, \citet{2018ApJ...854..174T} observed IRIS bombs with spatially-resolved bi-directional jets at the earliest stage of a flux emerging region, 
and found that they are associated with bald-patches and located in regions with large squashing factor at height of about 1\,Mm, 
which strongly suggested magnetic reconnection in the solar lower atmosphere.
Moreover, \citet{2018ApJ...856..127G} reported observations of long-lasting UV bursts occurring at the late phase of flux emergence while emerging flux tubes are interacting with ambient fields.
Therefore, activities in magnetic loops above flux emerging region could provide observational hints about how magnetic flux is emerging through and affecting the ambient field in the solar atmosphere.

\par
At the late phase of this flux emergence, very few energetic UV bursts were observed.
This is different from the behaviors of the flux emergence in the early phase.
Beside many loops with multiple temperatures that had been formed between the two major polarities,
we could also see smaller loops connecting the major polarities and the smaller ones.
These smaller loops could be formed by
(1) newly emerging flux that observed as small magnetic features with opposite polarity appearing at the side of and moving away from the major ones, 
and (2) preexisted loops that are dragged downward and observed as small magnetic features with the same polarity splitting and moving away from the major ones.
While the total magnetic flux is still increasing and the loops show systematic blue-shifts in the transition region Doppler map,
it indicates that the emerging flux should be still dominated at this stage.
These smaller loops might move toward and interact with each other, which leads to magnetic cancellation between small magnetic features.
A question is whether such interactions among the smaller loops could result in energy release heating the loops.
In the present observations, only brightenings in AIA 1700\,\AA\ are seen and no energetic event (such as UV bursts) has been detected in the IRIS \siiv\ spectral data.
This question remains open, and we will make a further investigation in a following-up work using IRIS sit-and-stare data that were taken after the raster data reported here.

\subsubsection{Heating of the loops}
We found that the nonthermal velocities of the loop system are generally smaller than the background in the transition region, but comparable in the corona.
This suggests that most of the emerging loops should be heated only after they reaching the transition region.
Because the footpoints of the loop system have higher temperatures and the flaring transition region loops are showing bright footpoints in prior,
it suggests that the heating processes should take place in the footpoints.

\par
The parameters (such as loop width, electron density and temperature profiles) of these loops measured with EIS data are similar to those in larger loops\,\citep[see e.g.][]{2005ApJ...633..499A,2008ApJ...686L.131W,2009ApJ...694.1256T,2009ApJ...700..762W}.
Actually, some larger loop systems studied previously also contain loop threads at temperatures from transition region to coronal\,\citep{2008ApJ...686L.131W,2009ApJ...694.1256T}.
Although those loop system might be experiencing similar flux emerging processes as the current one, the loop system observed here is much smaller (in length) and the heating therein could be much different.

\par
Not like many hot loops in the corona, the cool loops observed here show very dynamic evolution.
It suggests that they should be heated in an impulsive way.
We also observed that some flaring cool loops did not have any bright footpoints in the spectroscopic data.
This is consistent with recent one-dimensional simulations of loops with apex density more than $10^9$\,cm$^{-3}$ that were assumed to be heated by nano-flares occurring at the apex with heating either in an electron beam model with energy cutoff up to 15\,keV or in a thermal conduction model\,\citep{2018ApJ...856..178P}.
However, the loops observed here are systematically emerging and evolving and also they are much shorter than that modelled in \citet{2018ApJ...856..178P}.
Because heating mechanism also depends on loop length, whether the loops observed here were heated in the same way needs further modelling constrained by the present observations.
The following-up work using sit-and-stare data will also shed more light on this problem.

\par
The size of these loops is comparable to coronal bright points\,\citep[see][for a review of this topics]{madjarska_lrsp},
which also have multi-thermal nature\,\citep[see e.g.][]{2012A&A...545A..67M}.
One of the heating mechanisms of coronal bright points is believed to be magnetic reconnection among converging magnetic loops\,\citep[e.g.][]{1994ApJ...427..459P,2016ApJ...818....9M}.
Whether could such loop systems as observed here be formed and heated in the similar way?
This question is worthy to investigate further using higher-cadence IRIS data, and it will also help understand the possible heating mechanisms of coronal bright points.

\section{Conclusions}
\label{sect_conclusion}
In the present study, I report on IRIS, Hinode/EIS, Hinode/XRT, SDO/AIA and SDO/HMI observations of a magnetic loop system above a flux-emerging region.
The flux emergence is at its late phase that the two major polarities had been formed.
The separation between the two major polarity is about 12\arcsec.
At the side of a major polarity, small magnetic features with opposite polarity were emerging and moving toward the other major polarity.
Small magnetic features with the same polarity might also split from a major polarity and move toward the other major polarity.
The small magnetic features with opposite polarities might meet and cancel each other while they are moving toward.

\par
A set of loops had been formed between the two major polarities at the beginning of our observations.
Some of the loop threads are connecting the two major polarities and some others are connecting one major and one smaller polarities.
The cross-section of the loop threads is about 0.5\arcsec\ measured with IRIS SJ images.
We found that they consist of loop threads with temperatures from $2.5\times10^4$\,K (low transition region) to $2.8\times10^6$\,K (corona).
Most of loop threads with different temperatures are not identical in space suggesting that the loop threads appear to have temperatures in a small range.

\par
In the middle of the loop system, the Doppler maps show an upward velocity of $\sim$10\,\kms\  in the transition region (\siiv) and downward velocity of $\sim$10\,\kms\ in the corona (Fe\,{\sc xii}).
In the transition region, the nonthermal velocities of most of the loops are found to be less than 10\,\kms\ that are much smaller than the surrounding region.
While in the corona, they are not different from the surrounding region.
The electron densities of the loop system measured in coronal temperature are found to be in the range of 1--$4\times10^9$\,cm$^{-3}$ with larger values in the footpoints.
Using IRIS O\,{\sc iv} line pair, we are also able to measure electron density of  
a flaring loop thread in the transition region and found to be in the range of $2-8\times10^{10}$\,cm$^{-3}$ with an average of $\sim5\times10^{10}$\,cm$^{-3}$.
The DEM profiles derived with EIS iron lines in a few locations of the loops system imply that these locations should contain loop threads with different temperatures.

\par
Our observations indicate that the flux emergence in its late phase is much different from that at the early stage.
It might consist of magnetic reconnection between newly emerging flux and pre-existed flux, but it does not produce any UV burst that are signatures of magnetic reconnection in the lower solar atmosphere.
Most of the emerging loops at this stage are likely to be heated after they reaching the transition region or above because most of them have small nonthermal velocity in \siiv.
The dynamics of these loops suggests that they are heated impulsively,
and how it actually works requires further investigation.
In a following study, we will exploit an IRIS sit-and-stare dataset to investigate the evolution of this loop system, which will shed more light on their physics.

\acknowledgments
{\it Acknowledgments:}
I would like to thank the anonymous referee for the constructive comments and helpful suggestions, and Dr. Giulio Del Zanna for helpful discussions.
This research is supported by National Natural Science Foundation of China (41474150,  41627806, U1831112, 41404135).
Z.H. thanks the China Postdoctoral Science Foundation and the Young Scholar Program of Shandong University, Weihai (2017WHWLJH07).
The observation program at BBSO is supported by the Strategic Priority Research Program |
The Emergence of Cosmological Structures of the Chinese Academy of Sciences, Grant No.
XDB09000000.
Z.H. is grateful to BBSO, IRIS and Hinode operating teams for their help and to the BBSO staff for their hospitality while carrying out the observing campaign.
Z.H. acknowledges comments from Prof. Lidong Xia and useful discussion with Dr. Hui Tian.
IRIS is a NASA small explorer mission developed and operated by LMSAL with mission operations executed at NASA Ames Research center and major contributions to downlink communications funded by ESA and the Norwegian Space Centre.
Hinode is a Japanese mission developed and launched by ISAS/JAXA, collaborating with NAOJ as a domestic partner, NASA and STFC (UK) as international partners. Scientific operation of the Hinode mission is conducted by the Hinode science team organized at ISAS/JAXA. This team mainly consists of scientists from institutes in the partner countries. Support for the post-launch operation is provided by JAXA and NAOJ(Japan), STFC (U.K.), NASA, ESA, and NSC (Norway).
Courtesy of NASA/SDO, the AIA and HMI teams and JSOC.
CHIANTI is a collaborative project involving George Mason University, the University of Michigan (USA) and the University of Cambridge (UK).

\bibliographystyle{aasjournal}
\bibliography{bibliography}

\end{document}